%% file: main.tex
\documentclass[conference]{IEEEtran}
\IEEEoverridecommandlockouts
\usepackage{cite}
\usepackage{amsmath,amssymb,amsfonts}
\usepackage{graphicx}
\usepackage{textcomp}
\usepackage{xcolor}
\usepackage{booktabs} 
\usepackage{epsfig}
\usepackage{multicol}
\usepackage{algorithm}
\usepackage{algorithmic}
\usepackage{theorem}
\usepackage{wrapfig}
 
\newtheorem{example}{Example}[section]

\def\A{\mathbf{A}}
\def\B{\mathbf{B}}
\def\0{\mathbf{0}}
\def\M{\mathbf{M}}
\def\DDelta{\mathbf{\Delta}}
\def\PPsi{\mathbf{\Psi}}
\def\pp{\mathbf{\Psi}}

\def\firstelt{f}
\def\slope{s}
\def\compressed{\mathtt{C}}

\def\BibTeX{{\rm B\kern-.05em{\sc i\kern-.025em b}\kern-.08em
    T\kern-.1667em\lower.7ex\hbox{E}\kern-.125emX}}
\begin{document}

\title{Compressed Matrix Computations}

\author{\IEEEauthorblockN{Matthieu Martel}
\IEEEauthorblockA{\textit{LAMPS, University of Perpignan} \\
52, avenue Paul Alduy, Perpignan, France \\
matthieu.martel@univ-perp.fr}
}

\maketitle

\begin{abstract}
Frugal computing is becoming an important topic for environmental reasons. In this context, several techniques have been 
proposed  to reduce the storage of scientific data by dedicated compression methods specially tailored for arrays of floating-point numbers.
While these techniques are quite efficient to save memory, they introduce additional computations to compress and decompress the data
before processing them. In this article, we introduce a new lossy, fixed-rate compression technique for
2D-arrays of floating-point numbers which allows one to compute directly on the compressed data, without decompressing them.
We obtain important speedups since less operations are needed to compute among the compressed data and since no decompression and re-compression
is needed. More precisely, our technique makes it possible to perform basic linear algebra operations such as addition, multiplication
by a constant among compressed matrices and dot product and matrix multiplication among partly uncompressed matrices. 
This work has been implemented into a tool named blaz and a comparison with the
well-known compressor zfp in terms of execution-time and accuracy is presented.
\end{abstract}


\input{intro.tex}

\input{compscheme.tex}

\input{algos.tex}

\input{xp.tex}

\input{cc.tex}

\bibliographystyle{plain}
\bibliography{main}

\end{document}

%% file: intro.tex

\section{Introduction}

At the beginning of this new decade, more than ten percent of the world total electricity production is consumed by information and communication technology 
systems \cite{Jon18}. 
This huge consumption can be roughly decomposed into three equivalent parts: storage ($30\%$), computing ($30\%$) and networking ($40\%$).
This distribution imposes to who wants to propose frugal computing solutions, for ecological or maybe economical reasons, to attack these three points
all together: reducing storage, computing and networking simultaneously. In this article, we propose a solution to this problem for the special case of
matrix computations \cite{GL13} which is ubiquitous in computer science. 

Clearly, at the beginning of this new decade, scientific data remain the most important part
of the world stored data, before ERP or entertainment data  for example \cite{Hal19}. 
Then it urgent to seek ways of processing scientific data which offer simultaneously
gains in terms of storage, computing and networking.
In this article, the scientific data that we consider are  large matrices  of floating-point numbers \cite{IEEE754}
used to represent smooth curves (as in other inspiring work \cite{DC16,Lin14}, we will assume that there exist correlations between adjacent elements of matrices).

Because usual compression techniques \cite{HAR18,JTP21} are not efficient enough on scientific data, specific (lossy) compressors have been developed
such as \texttt{zfp} \cite{Lin14} or \texttt{sz} \cite{DC16}. The performance of these tools is impressive in terms of data size reduction and accuracy, 
allowing to reduce storage by $10$ to $20$ without significant loss of relevant information. 
Doing so, they help to reduce the footprint
of scientific data storage and communications. However, they increase the computing resources needed to solve a problem, since data must be 
uncompressed before processing and re-compressed after.  

In this article, we introduce a new lossy compression technique for scientific data which makes it possible to compute directly among the compressed data,
without decompression. While our method is not as efficient as \texttt{zfp} or \texttt{sz}  in terms of compression rate and accuracy, 
it allows one to compute directly with the 
compressed data. This avoids a huge amount of operations since $i)$ data do not need to be uncompressed and re-compressed and $ii)$ the compressed data
being smaller, less operations are needed to compute with them. More precisely, our technique operates upon two-dimensional arrays 
storing matrices and allows one to perform basic linear algebra operations such as addition and multiplication by a constant 
on the compressed data and dot product and matrix multiplication on partly decompressed matrices.  
To our knowledge, this is the first attempt to define a compression technique enabling one to compute directly on the
compressed data.

We can give a bird's eye view of our technique as follows. Our compressor  is lossy, fixed-rate (11.37 compression rate) and operates on $8\times 8$ blocks of IEEE754 \texttt{binary64} 
floating-point numbers \cite{IEEE754}. After a block-splitting stage, we perform successively stages of normalization, prediction, transformation and,
finally, quantization, all described in details in Section~\ref{seccompress}. 

A key-point is that all these stages act like linear maps (at the notable exception of the quantization done at the end of the scheme), 
i.e. the transformation $t$ applied at any stage (but quantization) of our scheme
satisfies $t(\A+\B)= t(\A)+t(\B)$  and $t(c\A)=c t(\A)$ where $\A$ and $\B$ are matrices represented in the adequate data structures
and where $c$ is a scalar constant. These properties are used in Section \ref{secalgo} to design algorithms to add compressed matrices and
to  multiply them by constants without decompression.
Additionally,  we will also see in Section \ref{secalgo} that the dot product among lines and columns of compressed matrices as well as the product $\A\times \B$
of matrices
can be
computed on partially uncompressed matrices. 

Last but not least, our compression scheme has been implemented into an open-source library named \texttt{blaz} and
experimental results are given in Section \ref{secxp}. They show that the loss of accuracy introduced by our technique  remains small, compared to the gains 
obtained in terms of execution time and storage. For instance, the addition of matrices of size $2000 \times 2000$  with \texttt{blaz} is more than $50$ times 
faster than with standard, uncompressed matrices and more than $5 000$ times faster than with \texttt{zfp} (see Section \ref{xpperf}). On the other hand,
The relative errors introduced by  \texttt{blaz} are less than $10$ times greater than with  \texttt{zfp} for our basis of examples (see Section \ref{xpacc}).  
This  corresponds to the loss of one more decimal digit than with \texttt{zfp}. We strongly believe that this compromise is quite acceptable
in many contexts.

As already mentioned, this article is organized as follows. The compression scheme used by \texttt{blaz} is detailed
in Section \ref{seccompress} and the algorithms used to compute among the compressed structures are introduced in Section \ref{secalgo}.
The experimental results are given in Section \ref{secxp} and Section \ref{seccc} concludes.

%% file: compscheme.tex
\section{Compression Scheme}
\label{seccompress}

\begin{figure}[t]
    \hrule
    \medskip
        \centerline{\includegraphics[width=0.95\columnwidth]{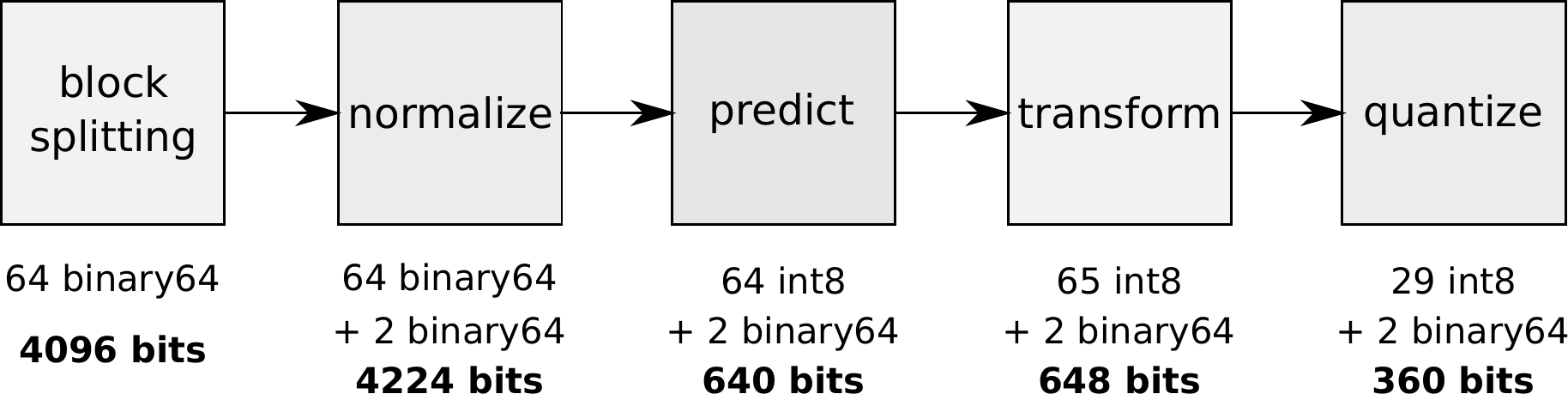}}
    \medskip
    \hrule
        \caption{\label{figscheme}Overview of \texttt{blaz} compression scheme.}
    \end{figure}

In this section, we introduce the five stages of our compression scheme which perform
block-splitting, normalization, prediction, transformation and quantization.
The originality of our compression scheme is to be compatible with the basic linear algebra operations that can be performed
directly on the compressed data structures using the algorithms introduced in Section \ref{secalgo}.
A summary of this scheme is given in Figure \ref{figscheme} and we detail the different steps hereafter.

\paragraph{Block Splitting}
First of all, as in most fixed-rate compressors \cite{Lin14}, we split the original matrix into blocks. In practice, our blocks have size $8\times 8$.

\begin{figure}[tb]\hrule
            \centerline{ \includegraphics[width=1.15\columnwidth]{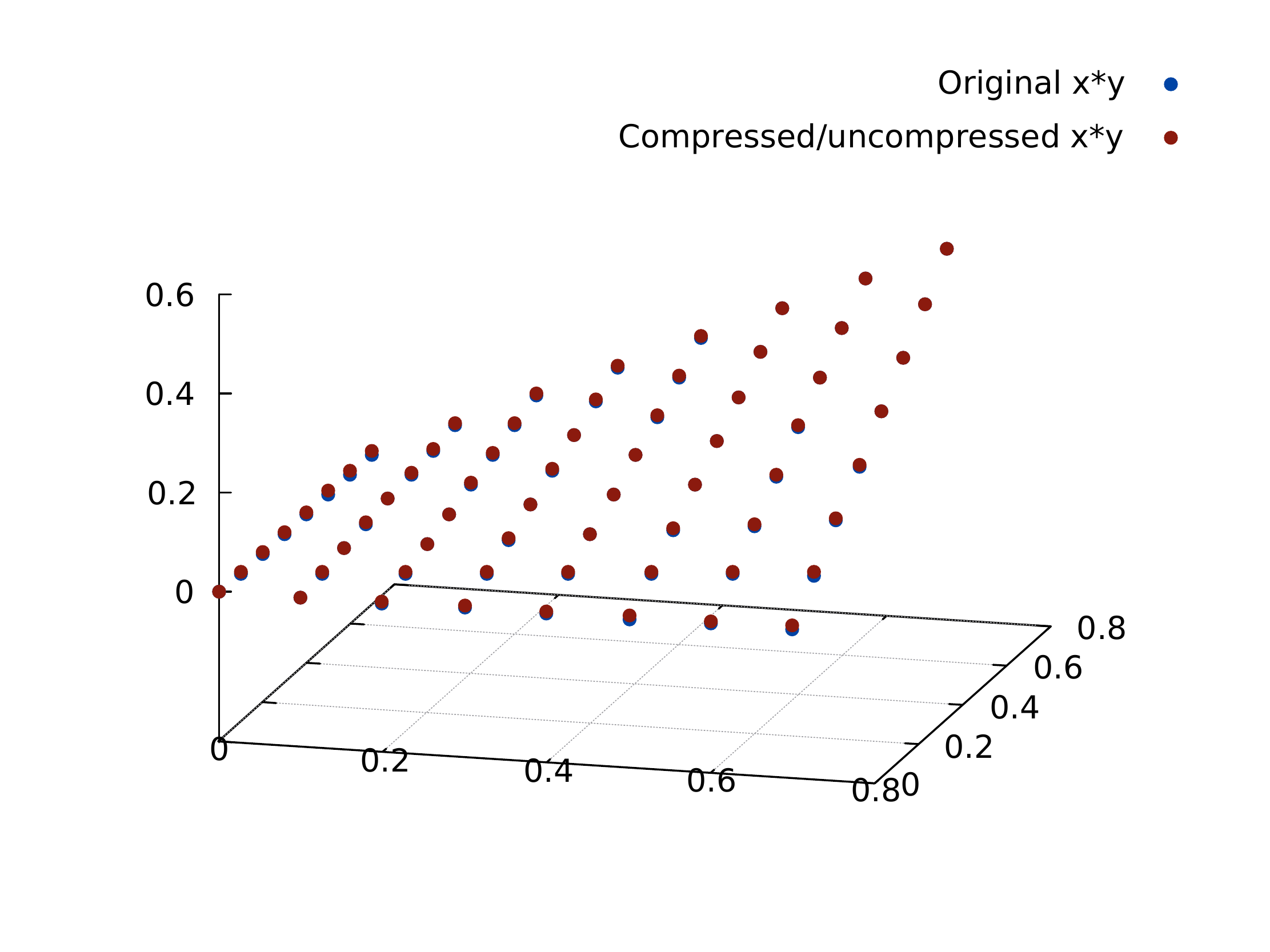}}
            \vspace{-0.6cm}
            \hrule
            \caption{\label{figxy}Comparison between the original block of Example \ref{ex1} and its compressed/uncompressed version.}
            \end{figure}

\begin{example}\label{ex1}
All along this section, we are going to illustrate how our compressor works using the following $8\times 8$ matrix  
\begin{equation*}
    \small
    \begin{array}{c}
        \A =\\ \\
 \left(\scriptsize\begin{array}{cccccccc}
    0.000 & 0.000 & 0.000 & 0.000 & 0.000 & 0.000 & 0.000 & 0.000 \\
    0.000 & 0.010 & 0.020 & 0.030 & 0.040 & 0.050 & 0.060 & 0.070 \\
    0.000 & 0.020 & 0.040 & 0.060 & 0.080 & 0.100 & 0.120 & 0.140 \\
    0.000 & 0.030 & 0.060 & 0.090 & 0.120 & 0.150 & 0.180 & 0.210 \\
    0.000 & 0.040 & 0.080 & 0.120 & 0.160 & 0.200 & 0.240 & 0.280 \\
    0.000 & 0.050 & 0.100 & 0.150 & 0.200 & 0.250 & 0.300 & 0.350 \\
    0.000 & 0.060 & 0.120 & 0.180 & 0.240 & 0.300 & 0.360 & 0.420 \\
    0.000 & 0.070 & 0.140 & 0.210 & 0.280 & 0.350 & 0.420 & 0.490 \\    
    \end{array}\normalsize\right).\end{array}
\end{equation*}
We may assume that $\A$  is one of the blocks obtained by splitting a larger  original matrix. 
The values of $\A$ correspond to the functions $f(x,y) = x\times y$ displayed in Figure \ref{figcurves} for $x$ and $y$ starting at $0$ with a step of $0.1$. 
The graphical representation of Block $\A$ is also given in Figure \ref{figxy}.  
\hfill$\blacksquare$
\end{example}

\paragraph{Block Normalization}
After block splitting, the next step of our scheme is to normalize the values. This stage is twofold. First, for each block element, we compute the difference between itself and 
its preceding elements. More precisely, taking as entry a $8\times 8$ matrix $\M$ corresponding to a block, we compute the new block $\DDelta$ such that, for any $0\le i,j\le 7$
\begin{equation}\label{eqnorm}\small
        \DDelta_{ij}=
\left\{
    \begin{array}{ll}
0&  i=j=0,\\     
\M_{0,j} - \M_{0,j-1} &  i=0,\ j\not=0,\\
\M_{i,0} - \M_{i-1,0} &  i\not=0,\ j=0,\\
\frac{\big(\M_{i,j} - \M_{i-1,j}\big) + \big(\M_{i,j} - \M_{i,j-1}\big)}{2}& \text{otherwise}. 
\end{array}\right.
\end{equation}
Normalization relies on the assumption that block elements are correlated and form a smooth surface. The differences between consecutive values are then assumed to be small
and this step aims at reducing the range of values occurring in the block. In addition to the matrix $\DDelta$, we have to store the value of the first element of the block $\M_{00}$
into a \texttt{binary64} number.

\begin{example}\label{ex2}
By normalizing the matrix $\A$ of Example \ref{ex1} using the formula displayed in Equation (\ref{eqnorm}), we obtain the new matrix 
    \begin{equation*}
        \small
        \begin{array}{c}
            \A_N =\\ \\    
     \left(\scriptsize\begin{array}{cccccccc}
        0.000 & 0.000 & 0.000 & 0.000 & 0.000 & 0.000 & 0.000 & 0.000 \\
        0.000 & 0.010 & 0.015 & 0.020 & 0.025 & 0.030 & 0.035 & 0.040 \\
        0.000 & 0.015 & 0.020 & 0.025 & 0.030 & 0.035 & 0.040 & 0.045 \\
        0.000 & 0.020 & 0.025 & 0.030 & 0.035 & 0.040 & 0.045 & 0.050 \\
        0.000 & 0.025 & 0.030 & 0.035 & 0.040 & 0.045 & 0.050 & 0.055 \\
        0.000 & 0.030 & 0.035 & 0.040 & 0.045 & 0.050 & 0.055 & 0.060 \\
        0.000 & 0.035 & 0.040 & 0.045 & 0.050 & 0.055 & 0.060 & 0.065 \\
        0.000 & 0.040 & 0.045 & 0.050 & 0.055 & 0.060 & 0.065 & 0.070 \\        
        \end{array}\normalsize\right).\end{array}
    \end{equation*}
Let us observe that the values in $\A_N$ are smaller than those in $\A$. We also store the element $\A_{{00}}= 0.0$ into a separate \texttt{binary64} number. \hfill$\blacksquare$
    \end{example}

The second step of the normalization stage consists of dividing the values in $\DDelta$ by the mean 
 slope between consecutive values of the block $\A$ (considering only the non zero elements.)
Since $\DDelta$ already contains the differences between adjacent elements of $\A$, we have
\begin{equation}\label{eqslope}\small
s = \frac{1}{K}\sum_{0\le i,j\le 7} |\DDelta_{ij}|
\end{equation}
with
\begin{equation}\label{eqslope2}\small
  \ K=\text{Card}\big(\{\DDelta_{ij}, \ 0\le i,j\le 7\ : \DDelta_{ij}\not= 0)\}\big)
    \end{equation}
and we compute
\begin{equation}\label{eqslope2}\small
\DDelta' = \frac{1}{s}\cdot\DDelta
    \end{equation}
Note that the mean slope $s$ must be stored into a \texttt{binary64} number by our compressor.
        
\begin{example}\label{ex3}
Using the matrix $\A_N$ of Example \ref{ex2}, we have $s=0.04$ and the new matrix $\A'_N = \frac{1}{s}\cdot \A_N$ is valuated to
    \begin{equation*}
        \small\begin{array}{c}
    \A'_N =\\ \\
    \left(\scriptsize\begin{array}{cccccccc}
        0.000 & 0.000 & 0.000 & 0.000 & 0.000 & 0.000 & 0.000 & 0.000 \\
        0.000 & 0.250 & 0.375 & 0.500 & 0.625 & 0.750 & 0.875 & 1.000 \\
        0.000 & 0.375 & 0.500 & 0.625 & 0.750 & 0.875 & 1.000 & 1.125 \\
        0.000 & 0.500 & 0.625 & 0.750 & 0.875 & 1.000 & 1.125 & 1.250 \\
        0.000 & 0.625 & 0.750 & 0.875 & 1.000 & 1.125 & 1.250 & 1.375 \\
        0.000 & 0.750 & 0.875 & 1.000 & 1.125 & 1.250 & 1.375 & 1.500 \\
        0.000 & 0.875 & 1.000 & 1.125 & 1.250 & 1.375 & 1.500 & 1.625 \\
        0.000 & 1.000 & 1.125 & 1.250 & 1.375 & 1.500 & 1.625 & 1.750 \\        
    \end{array}\normalsize\right).\end{array}
    \end{equation*}
Note that in $\A'_N$, for the sake of clarity, the values are rounded to the nearest after three decimal digits. Obviously, these values are computed in IEEE754 double precision in our
implementation. \hfill$\blacksquare$
    \end{example}

\paragraph{Prediction}\label{ssecpredict}
This third stage of our compression scheme consists of replacing the slopes of the block $\DDelta'$ by a prediction. 
\begin{wrapfigure}[]{r}{4cm}
    \begin{minipage}{4cm}
        \hrule
        \medskip
            \centerline{\includegraphics[width=4cm]{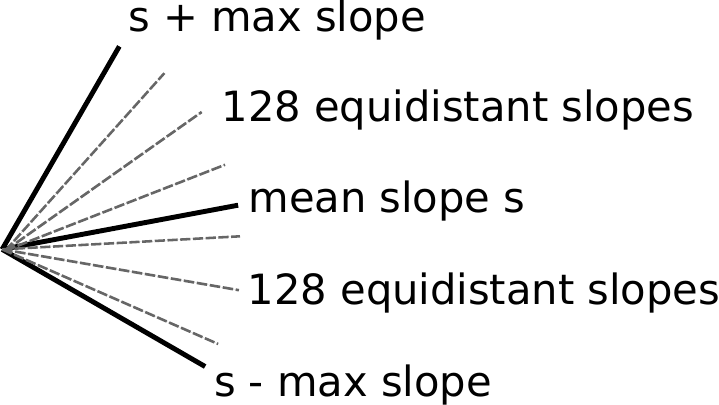}}
        \medskip
        \hrule
    \end{minipage}
            \caption{\label{figslope}Slopes of reference used in our prediction.}
    \end{wrapfigure}            
    Let $s_{max}$ be the maximal value in $\DDelta'$ in absolute value,
$s_{max}=\max\{|\DDelta'_{ij}|, \ 0\le i,j\le 7\}$, we take a set $S=\{s_0,s_1,\ldots,s_{255}\}$ of $256$ equidistant points in the 
interval $[s-s_{max},s+s_{max}]$ as our slopes of 
reference (see Figure \ref{figslope}). Then each value in $\DDelta'$ is approximated by its closest element in $S$. 
Assuming that, for all $0\le k <255$, $s_k\le s_{k+1}$, we store in a new matrix $\PPsi$ the indexes $k$ of the slopes $s_k$ which are the best approximations of the values in $\DDelta'$.
Formally, for $0\le i,j<8$, we have
\begin{equation}\label{eqpredict}\small
\PPsi_{ij} = k \ \text{such that}\ |\DDelta_{ij}-s_k|\le |\DDelta_{ij}-s_\ell|,\ \forall 0\le \ell\le 255. 
\end{equation}
Note that $\PPsi$ is a $8\times 8$ matrix of integers encoded on 8 bits. In addition to the mean slope $s$, we need to store $s_{max}$ into a \texttt{binary64} floating-point number.

\begin{example}\label{ex4}
    Using the matrix $\A_N'$ of Example \ref{ex3}, we compute the prediction matrix 
        \begin{equation*}
            \small
            \begin{array}{c}
                \mathbf{P} =\\ \\        
          \left(\scriptsize\begin{array}{cccccccc}
            125  & 125  & 125  & 125  & 125  & 125  & 125  & 125  \\
125  & 143  & 153  & 162  & 171  & 180  & 189  & 198  \\
125  & 153  & 162  & 171  & 180  & 189  & 198  & 207  \\
125  & 162  & 171  & 180  & 189  & 198  & 207  & 217  \\
125  & 171  & 180  & 189  & 198  & 207  & 217  & 226  \\
125  & 180  & 189  & 198  & 207  & 217  & 226  & 235  \\
125  & 189  & 198  & 207  & 217  & 226  & 235  & 244  \\
125  & 198  & 207  & 217  & 226  & 235  & 244  & 253  \\
        \end{array}\normalsize\right).\end{array}
        \end{equation*}
The \texttt{binary64} values $A_{00}=0.0$, and $s=0.04$  are also stored.
        \hfill$\blacksquare$
        \end{example}

\paragraph{Block Transform and Quantization}
As in many other compressors (e.g. JPEG-2000 \cite{HAR18}), we  use a Type II two-dimensional discrete cosine transform (DCT, \cite{Str99})  of the block resulting from the normalization stage.
DCTs are used to aggregate large coefficients in the first lines and columns of a matrix, small values occurring in the other elements after transformation.  
The quantization then consists of considering that these small coefficients are equal to $0$ and of avoiding to store them in the compressed matrix. Our compressor only keeps the
elements of the first two lines and columns of the matrix, i.e. $28$ values. Note that the values returned by the DCT may be larger than $127$ in absolute value. 
Let $m$ be the greatest value in absolute value of the matrix $\mathbf{D}$ resulting from the transform.
We re-scale $\mathbf{D}$ by multiplying all its elements by 
\begin{equation}\label{eqphi}
\varphi=\frac{127}{m}. 
\end{equation}
In order to recover the original matrix during the decompression phase, the value $\varphi$ must also be 
stored into a $8$ bits integer.

\begin{example}\label{ex5}
After re-scaling, the DCT of matrix $\mathbf{P}$ of Example \ref{ex4} gives the new matrix
        \begin{equation*}\small
        \mathbf{P'} = \left(\scriptsize\begin{array}{cccccccc}
            \mathbf{122} & \mathbf{-51} & \mathbf{-11} & \mathbf{-14} & \mathbf{-8} & \mathbf{-8} & \mathbf{-4} & \mathbf{-2}  \\
            \mathbf{-51} & \mathbf{15} & \mathbf{7} & \mathbf{7} & \mathbf{5} & \mathbf{4} & \mathbf{3} & \mathbf{1}  \\
            \mathbf{-11} & \mathbf{7} & 0 & 0 & 0 & 0 & 0 & 0  \\
            \mathbf{-14} & \mathbf{7} & 0 & 1 & 0 & 0 & 0 & 0  \\
            \mathbf{-8} & \mathbf{5} & 0 & 0 & 0 & 0 & 0 & 0  \\
            \mathbf{-8} & \mathbf{4} & 0 & 0 & 0 & 0 & 0 & 0  \\
            \mathbf{-4} & \mathbf{3} & 0 & 0 & 0 & 0 & 0 & 0  \\
            \mathbf{-2} & \mathbf{1} & 0 & 0 & 0 & 0 & 0 & 0  \\            
        \end{array}\normalsize\right).
        \end{equation*}
For this example, all the non-zero coefficient but one in position $(4,4)$ already occur in the first two lines and columns of the matrix $\mathbf{P}'$ (in bold in the equation above) 
and the quantization only sets $\mathbf{P}'_{44}$ to $0$ introducing a slight information loss.
Recall that, in addition to $\mathbf{P}'$, the values $A_{00}=0.0$, $s=0.04$ and the scale factor $\varphi=20$ are also stored.
        \hfill$\blacksquare$
        \end{example}

\begin{example}
The difference between the original block of Example \ref{ex1} and the result of its compression and decompression by our method is displayed in
Figure \ref{figxy}.
\hfill$\blacksquare$
\end{example}

Following our scheme, a compressed block is stored using $29$ $8$-bits integers and $2$ \texttt{binary64} floating-point numbers.
The $29$ $8$-bits integers correspond to the $28$ values that we keep after quantization plus one $8$-bits integer corresponding to the re-scaling factor $\varphi$.
A block is then stored into
$360$ bits instead of the original $4096$ bits needed to store $64$   \texttt{binary64} numbers, yielding a compression rate of $11.37$.

%% file: algos.tex
\section{Basic Linear Algebra Algorithms}
\label{secalgo}

In this section, we introduce the basic linear algebra algorithms used to compute among matrices
following the scheme of Section \ref{seccompress}. More precisely, in what follows, we are going to
introduce successively the algorithms for addition, multiplication by a constant, dot product and matrix multiplication. 
These operations are detailed for elementary $8\times 8$ blocks. Obviously, for larger matrices, this has to be repeated for each elementary block.

All along this section, we use the following notations. Let $\mathbf{B}$ denote a $8\times 8$  block of a compressed matrix, $f$
and $s$ denote  respectively the values of the first element $\mathbf{B}_{00}$ and of the  mean slope $s$ of $\mathbf{B}$ as defined in Equation (\ref{eqslope}).
Both $f$ and $s$ are \texttt{binary64} numbers.
Next, $\varphi$ denotes the scale factor $\varphi$ introduced in Equation (\ref{eqphi}) to normalize the result of the DCT.
Recall that $\varphi$ is a $8$-bits integer.
Finally, $\compressed$ is an array of $28$ $8$-bits values containing the coefficients quantized after the DCT. 
\paragraph{Addition}
The algorithm for the addition $\mathbf{B}=\mathbf{B}_1+\mathbf{B}_2$ of two compressed blocks  is given in Algorithm \ref{algoadd}.
First of all, at Line $1$, the first elements $f$ of $\mathbf{B}$ is the addition $f_1+f_2$ of the first elements of $\mathbf{B}_1$ and $\mathbf{B}_2$. 
Similarly,  the mean slope of $\mathbf{B}$ is defined by $s=s_1+s_2$. 
Next, we want $\varphi=\frac{127}{m}$ with $m=m_1+m_2$. Then we have
\begin{equation}\small\label{eq55}
\varphi=\frac{127}{m_1+m_2}= \frac{127}{\frac{127}{\varphi_1}+\frac{127}{\varphi_2}} = \frac{1}{\frac{1}{\varphi_1}+\frac{1}{\varphi_2}} = \frac{\varphi_1 \varphi_2}{\varphi_1+\varphi_2}. 
\end{equation}

Equation (\ref{eq55}) motivates  the computation carried out at Line $2$ of Algorithm \ref{algoadd}.
Let $\mathbf{D}_1$ and $\mathbf{D}_2$ denoted the blocks obtained after the DCT by passing $\mathbf{B}_1$ and $\mathbf{B}_2$ through the scheme of Figure \ref{figscheme}.
Intuitively, if no re-scaling were done in our scheme, we could simply add $\mathbf{D}_1$ and $\mathbf{D}_2$ to obtain the block $\mathbf{D}$ corresponding to $\mathbf{B}$.
But two re-scaling are done in our scheme, during the normalization stage and among the values resulting from the DCT.
The coefficients $\alpha_1$ and $\alpha_2$ are used to transform blocks made for scale factors $\varphi_1$ and $\varphi_2$ to the new scale factor $\varphi$, accordingly to
Equation (\ref{eqslope2}). Similarly, $\beta_1$ and $\beta_2$ are used to adapt the scale of the blocks resulting from the DCT to the scale factor $\varphi$. 
It is then possible to re-scale and add the coefficients of the DCT contained in $\compressed_1$ and $\compressed_2$ as done at Line~$6$, in the for loop of Algorithm \ref{algoadd}.
Note that adding the coefficients is possible since the DCT defines a linear map among our blocks. Formally, for a $8\times 8$ block
$\pp$ as defined in Equation (\ref{eqpredict}), the two-dimensional DCT of $\pp$, denoted $\text{DCT}(\pp)=\mathbf{D}$ is defined, for $0\le i,j < 8$,  by
\begin{equation}
\small\label{eqdct}
\mathbf{D}_{ij} = \alpha_i \alpha_j \sum_{u=0}^7\sum_{v=0}^7 \pp_{ij} \cos\left[ \frac{(2u+1)i\pi}{16}\right] \cos\left[ \frac{(2v+1)j\pi}{16}\right]
\end{equation}
and it follows that for two blocks $\pp''=\pp+\pp'$,
\begin{equation}
    \label{eqdct2}
    {\scriptsize
\begin{array}{cl}
   & \mathbf{D}_{ij}+\mathbf{D'}_{ij} 
    \\
    \\=& \alpha_i \alpha_j \sum\limits_{u=0}^7\sum\limits_{v=0}^7 \pp_{ij} \cos\left[ \frac{(2u+1)i\pi}{16}\right] \cos\left[ \frac{(2v+1)j\pi}{16}\right]\\
\\
+ &\alpha_i \alpha_j \sum\limits_{u=0}^7\sum\limits_{v=0}^7 \pp'_{ij} \cos\left[ \frac{(2u+1)i\pi}{16}\right] \cos\left[ \frac{(2v+1)j\pi}{16}\right]\\
\\
= &\alpha_i \alpha_j \sum\limits_{u=0}^7\sum\limits_{v=0}^7 (\pp_{ij}+\pp'_{ij}) \cos\left[ \frac{(2u+1)i\pi}{16}\right] \cos\left[ \frac{(2v+1)j\pi}{16}\right]\\
&\\
= &\mathbf{D}''_{ij},\ 0\le i,j,<8
\end{array}}
\end{equation}
where the coefficients $\alpha_i$ are the usual DCT coefficients  defined by
\begin{equation}\small
    \alpha_0=\sqrt{\frac{1}{8}}\quad \text{and}\quad \alpha_i=\sqrt{\frac{1}{4}},\ 1\le i<8\enspace .
\end{equation}

 \begin{example}
    Figure \ref{figadd} displays the points of the result of the addition of two $8\times 8$ blocks corresponding to the functions $x\times y$ and $x^2\times y^2$
    for $x$ and $y$ starting at $(0,0)$ with a step of $0.1$. We display the result of the addition between the uncompressed blocks and the
    result of the addition of the compressed blocks (this latter result being uncompressed after the operation). \hfill $\blacksquare$
 \end{example}

Let us remark that the subtraction of blocks works similarly to the addition. We omit to detail it in the present article.

\begin{algorithm}[b]
    \caption{\label{algoadd}Addition of two compressed matrix blocks $\mathbf{B}_1$ and $\mathbf{B}_2$.}
    \begin{algorithmic}[1]\small 
    \STATE $\firstelt \leftarrow \firstelt_1 + \firstelt_2$ ; $\slope \leftarrow \slope_1 + \slope_2$
    \STATE $\varphi \leftarrow $int$_8\big((\varphi_1 \times \varphi_2)
                                                        \div (\varphi_1 + \varphi_2)\big)$
    \STATE $\alpha_1 \leftarrow \slope_1 \div (\slope_1 + \slope_2)$ ; $\alpha_2 \leftarrow \slope_2 \div (\slope_1 + \slope_2)$
    \STATE $\beta_1 \leftarrow \alpha_1 \div \varphi_1 \times \varphi$ ; $\beta_2 \leftarrow \alpha_2 \div \varphi_2 \times \varphi$
    \FOR{$i\leftarrow 0$ to $27$}
    \STATE $\compressed[i] \leftarrow $int$_8(\beta_1 \times \compressed_1[i] + \beta_2 \times \compressed_2[i])$
    \ENDFOR
    \RETURN $\firstelt$, $\slope$, $\varphi$, $\compressed$
    \end{algorithmic}
    \end{algorithm}

\paragraph{Multiplication by a Constant} In our framework, the multiplication by a constant $c$ is the simplest operation to implement.
The first element of the block and the slope are multiplied by $c$. The other elements remain unchanged. 
This is summarized in Algorithm \ref{algomulcst}. This algorithm is straightforward and we do not give more details about it.

\begin{algorithm}[b]
    \caption{\label{algomulcst}Multiplication of a compressed matrix block $\mathbf{B}_1$ by a constant $c$.}
    \begin{algorithmic}[1]\small 
    \STATE $\firstelt \leftarrow \firstelt_1 \times c$ ; $\slope \leftarrow \slope_1 \times c$
    \STATE $\varphi \leftarrow \varphi_1 $
    \FOR{$i\leftarrow 0$ to $27$}
    \STATE $\compressed[i] \leftarrow \compressed_1[i] $
    \ENDFOR
    \RETURN $\firstelt$, $\slope$, $\varphi$, $\compressed$
    \end{algorithmic}
    \end{algorithm}

\paragraph{Dot product}
Let $\mathbf{B}_1$ and $\mathbf{B}_2$ be two $8\times 8$ matrix blocks and let $\langle \mathbf{B}_1, \mathbf{B}_2\rangle_{ij}$ denote the dot product between the
$i^{th}$ line of $\mathbf{B}_1$ and the $j^{th}$ column of $\mathbf{B}_2$, $0\le i,j<8$. Our dot product requires a partial decompression of the blocks. More precisely,
we have to recover the values resulting from the prediction stage of our  scheme (third step of Figure \ref{figscheme} and Section \ref{seccompress}-c) 
and this implies to compute the inverse discrete cosine transform (IDCT) of the values contained in 
$\mathtt{C}_1$ and $\mathtt{C}_2$ (obviously, the values discarded by the quantization are replaced by zeros.) 

Let $\mathbf{P}_1=\mathtt{IDCT}(\mathtt{C}_1)$ and $\mathbf{P}_2=\mathtt{IDCT}(\mathtt{C}_2)$ be the blocks obtained by applying the inverse discrete cosine transform to $\mathtt{C}_1$
and $\mathtt{C}_2$. We have to compute
\begin{equation}\label{eqbbb0}\small
    \langle \mathbf{B}_1, \mathbf{B}_2\rangle_{ij} = \sum_{k=0}^7 \mathbf{B}_{1_{ik}} \mathbf{B}_{2_{kj}}.
\end{equation}
where the coefficients of $\mathbf{B}_{\ell}$,  $\ell\in \{1,2\}$, are defined by
\begin{equation}\label{eqbbb}\small
    \mathbf{B}_{\ell_{ij}} = 
    \left\{\begin{array}{l}
        f_\ell\ \text{if} \ i=j=0\enspace , \\
        \\
        \mathbf{B}_{\ell_{0,j-1}} + s_\ell\cdot\mathbf{P}_{\ell_{0,j-1}}, \ \text{if} \ i=0,\ 1\le j\le 7\enspace , \\
        \\
        \mathbf{B}_{\ell_{i-1,0}} + s_\ell\cdot\mathbf{P}_{\ell_{i-1,0}}, \ \text{if} \ 1\le i\le 7,\ j=0\enspace , \\
        \\
        \frac{\mathbf{B}_{\ell_{i-1,j}}+ \mathbf{B}_{\ell_{i,j-1}}}{2} + s_\ell\cdot\mathbf{P}_{\ell_{i-1,j-1}}    \ \text{otherwise} \enspace.
    \end{array}\right.
\end{equation}
Note that we do not need to compute the entire blocks $\mathbf{B}_1$ and $\mathbf{B}_2$. It is sufficient to compute the first $i$ lines of
$\mathbf{B}_1$ and the first $j$ columns of $\mathbf{B}_2$ in order to compute the dot product $ \langle \mathbf{B}_1, \mathbf{B}_2\rangle_{ij}$.
Algorithm \ref{algodp} summarizes how we perform the dot product following equations (\ref{eqbbb0}) and (\ref{eqbbb}).
In this algorithm, the function $\mathtt{IPREDICT\_LINES}(\mathtt{C},i)$ and $\mathtt{IPREDICT\_COLS}(\mathtt{C},j)$
use Equation (\ref{eqbbb}) to compute respectively the first $i$ lines and $j$ columns of $\mathbf{B}_1$ and $\mathbf{B}_2$.

\begin{algorithm}[t]
    \caption{\label{algodp}Dot product between Line $i$ and Column $j$ of the compressed matrix blocks $\mathbf{B}_1$ and $\mathbf{B}_2$.}
    \begin{algorithmic}[1]\small 
        \STATE $r \leftarrow 0$   

        \STATE $\mathbf{P}_1 \leftarrow \mathtt{IDCT}(\mathtt{C}_1)$ ; $\mathbf{P}_2 \leftarrow \mathtt{IDCT}(\mathtt{C}_2)$
        \STATE $\mathbf{B}_1 \leftarrow \mathtt{IPREDICT\_LINES}(\mathtt{P}_1,i)$
        \STATE$\mathbf{B}_2 \leftarrow \mathtt{IPREDICT\_COLS}(\mathtt{P}_2,j)$

        \FOR{$k\leftarrow 0$ to $7$}
            \STATE $r \leftarrow r + \mathbf{B}_{1}[i,k] \times \mathbf{B}_{2}[k,j] $
    \ENDFOR
    \RETURN $r$
    \end{algorithmic}
    \end{algorithm}

\paragraph{Matrix multiplication} Our algorithm for the multiplication $\mathbf{B}=\mathbf{B}_1\times\mathbf{B}_2$ of two blocks
uses the same ideas than for the dot product
and we do not detail it hereafter. The difference comes from the fact that the blocks resulting from the IDCT and from Equation
(\ref{eqbbb}) are needed many times since, for each $0\le i,j\le 7$, the element $\mathbf{B}_{ij}$ corresponds to the dot product
$\mathbf{B}_{ij} = \langle \mathbf{B}_{1}, \mathbf{B}_{2}\rangle_{ij}$. 
To avoid redundant computations, these blocks are computed only once.

For large matrices made of many $8\times 8$ blocks, the algorithms introduced in this section need to generalized. 
Briefly speaking, this corresponds to apply block-wise the former algorithms.

%% file: xp.tex
\section{Experimental Evaluation}
\label{secxp}

Our compression scheme as well as the operations among compressed matrices have been implemented into an open-source library
named \texttt{blaz}\footnote{\texttt{https://github.com/mmartel66/blaz}}, written in \texttt{C}, and,
in this section, we present two kinds of experiments, concerning the time performances  (Section \ref{xpperf}) and the accuracy (\ref{xpacc})
of \texttt{blaz}.
Altogether these experiments show that our technique makes it possible to save simultaneously storage and computation time. 
In both cases, comparisons with the \texttt{zfp} \cite{Lin14} library are reported. All these experiments have been carried out on a
quad core Intel Core i7-1165G7 processor with 16 gigabytes of RAM and running Ubuntu 20.04.3 LTS.
For the sake of simplicity, all our experiments are carried out on square matrices even though our techniques works for matrices of any dimensions.
For the relevance of the comparison, the compression rate of \texttt{zfp} is set to a ratio equivalent to \texttt{blaz}.

\begin{figure}[b]
    \hrule
    \vspace{0.2cm}
    \centerline{
    \includegraphics[width=\columnwidth]{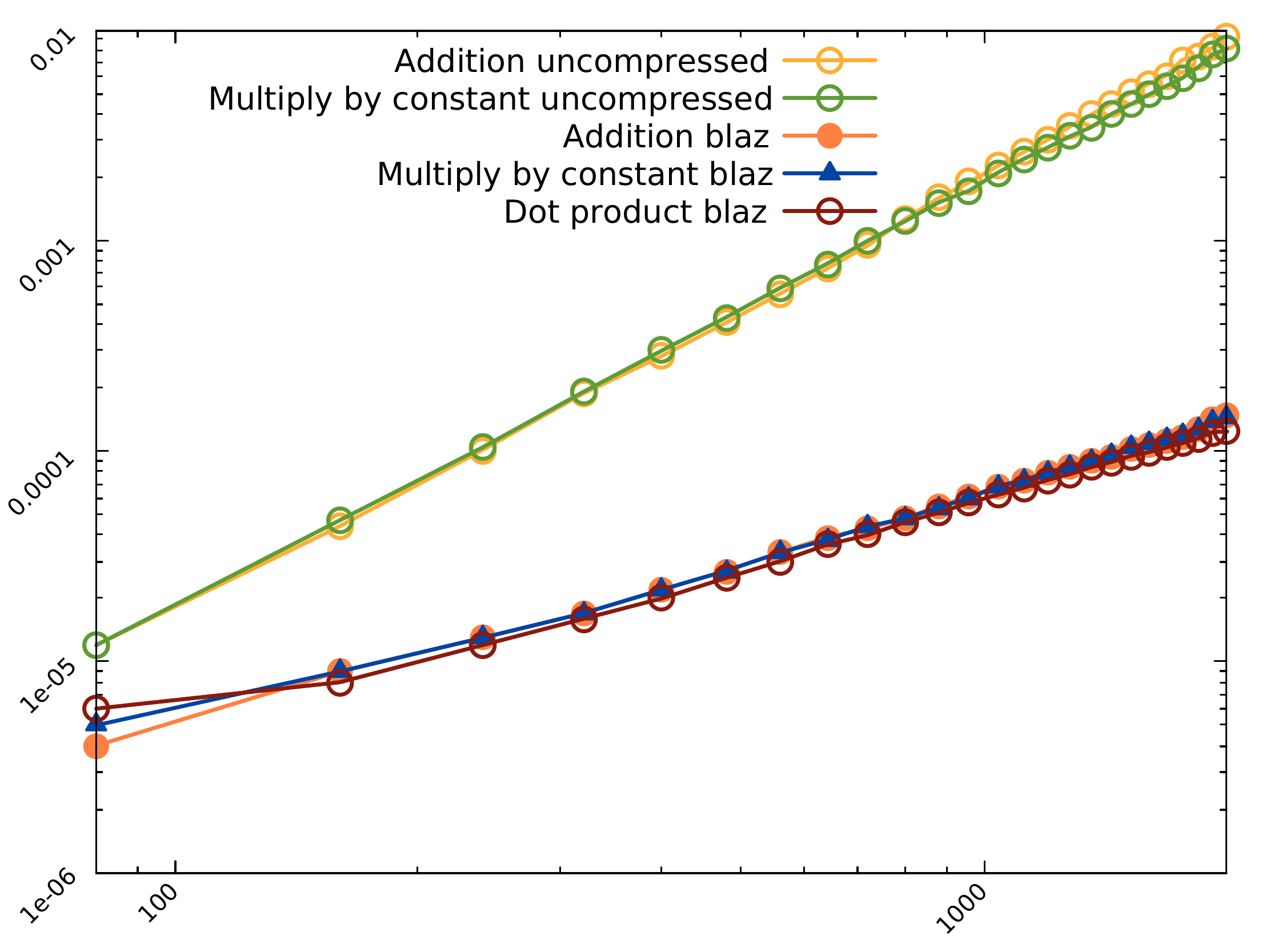}
    }
    \vspace{0.2cm}
    \hrule
    \caption{\label{fighs}Time measurement of operations in function of the size of the matrices for addition,
    multiplication by constant and dot product.}
\end{figure}

\begin{figure}[t]
    \hrule
    \vspace{0.2cm}
    \centerline{
    \includegraphics[width=\columnwidth]{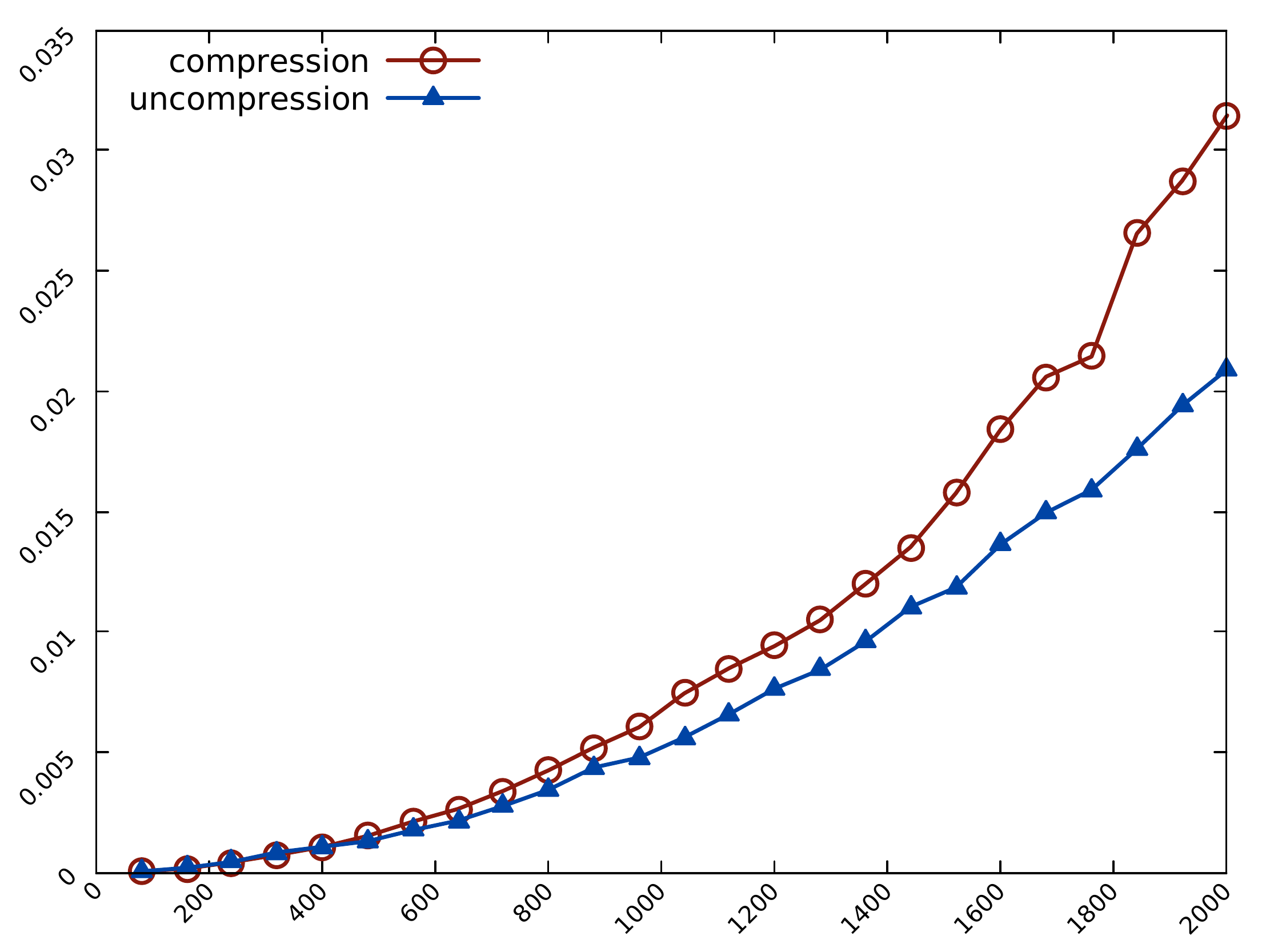}
    }
    \vspace{0.2cm}
    \hrule
    \caption{\label{figcu}Time measurement of compression and decompression operations in function of the size of the matrices.}
\end{figure}

\begin{figure*}[t]
    \hrule
    \vspace{0.2cm}
    \centerline{
    \begin{tabular}{ccc}
    \includegraphics[width=0.47\textwidth]{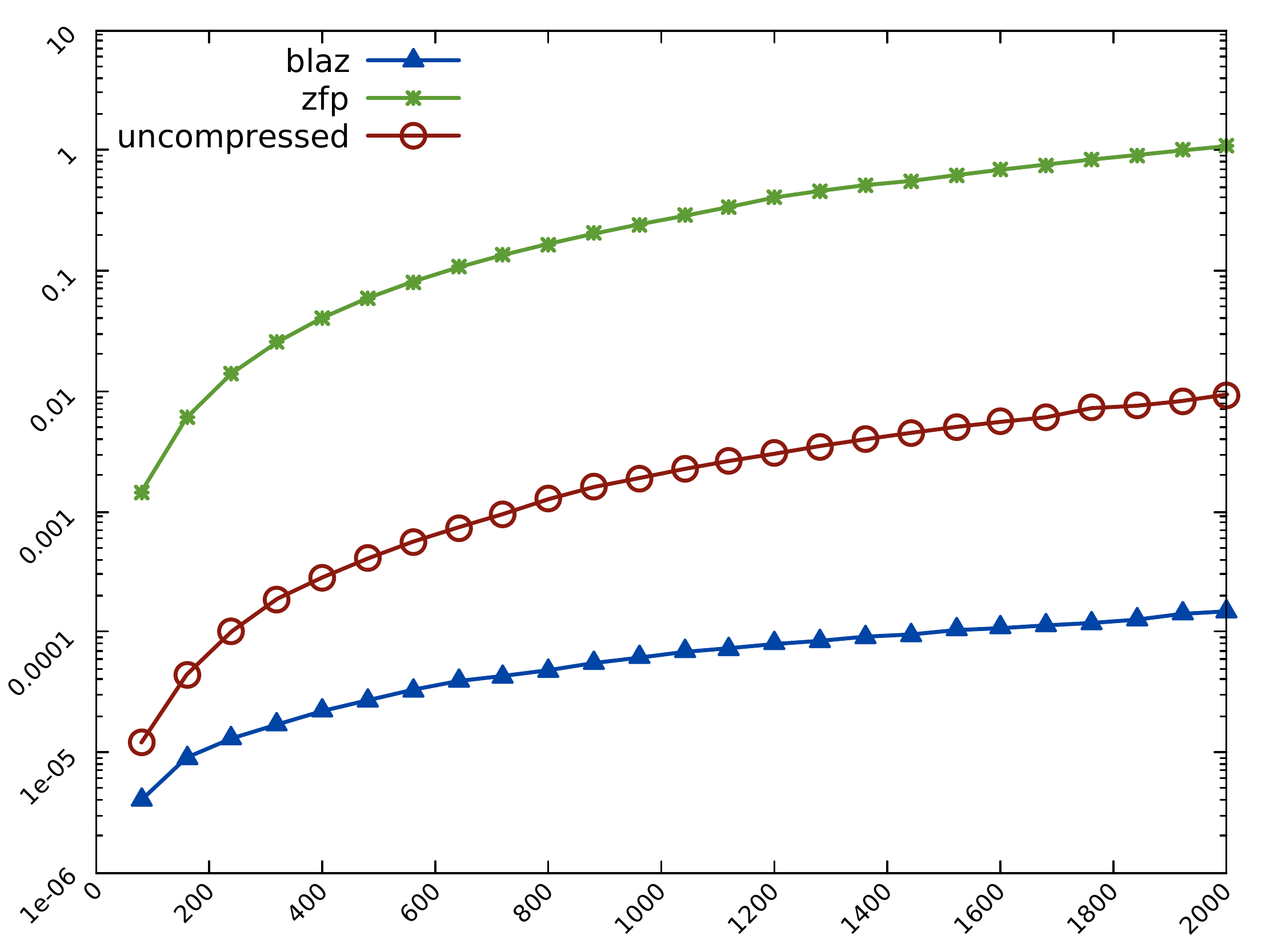}        && \includegraphics[width=0.47\textwidth]{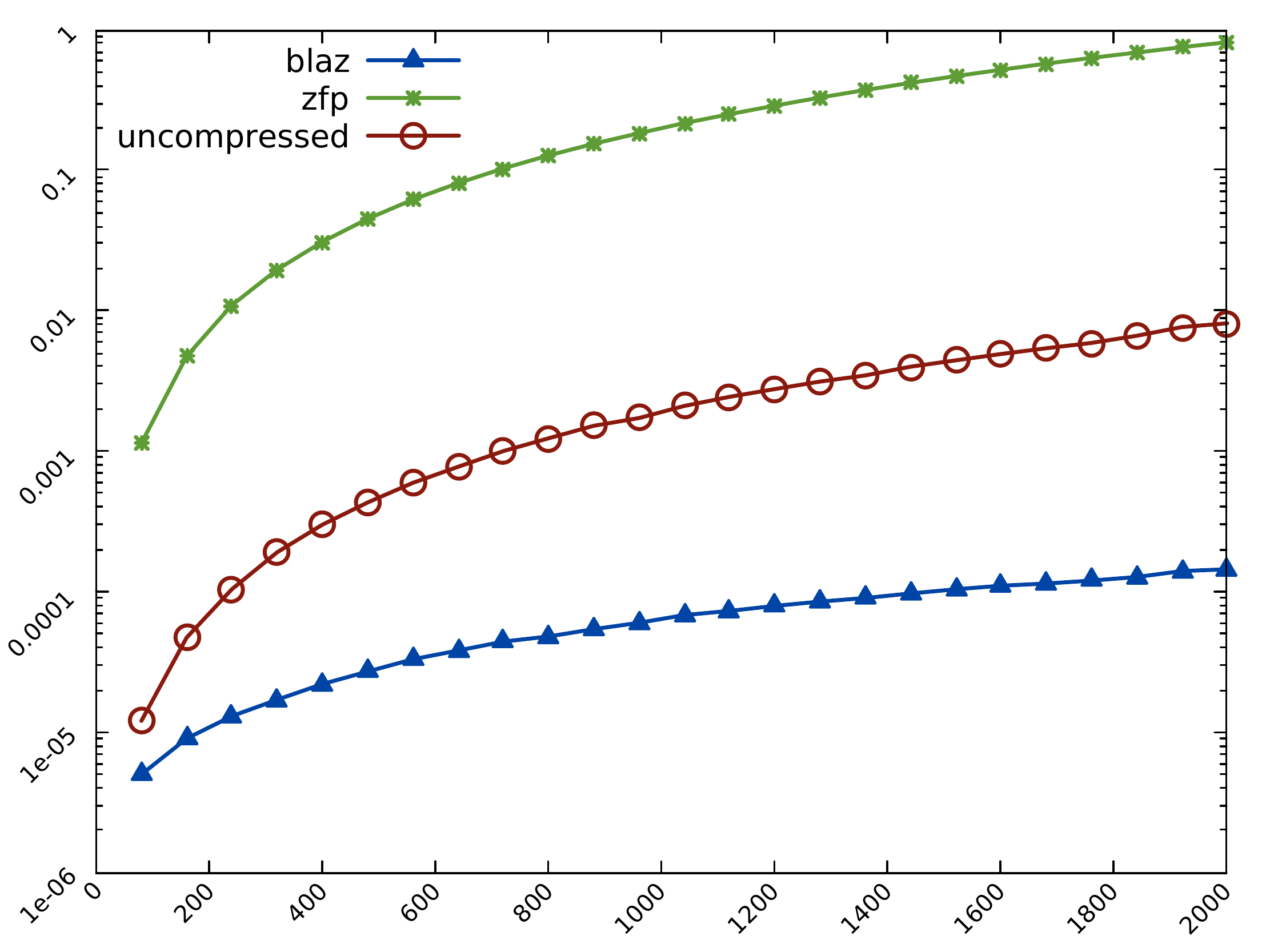}\\
    \includegraphics[width=0.47\textwidth]{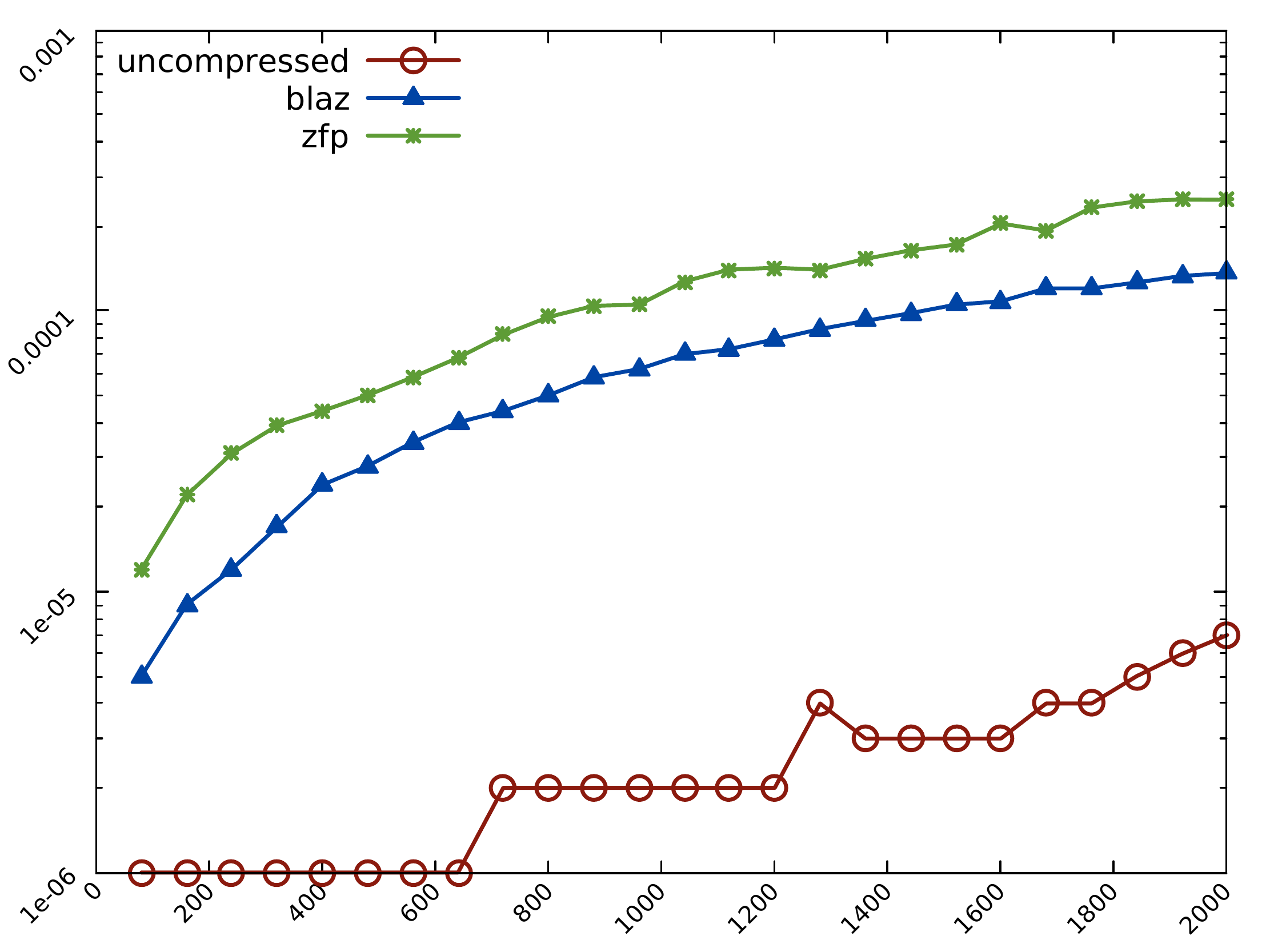}&        & \includegraphics[width=0.47\textwidth]{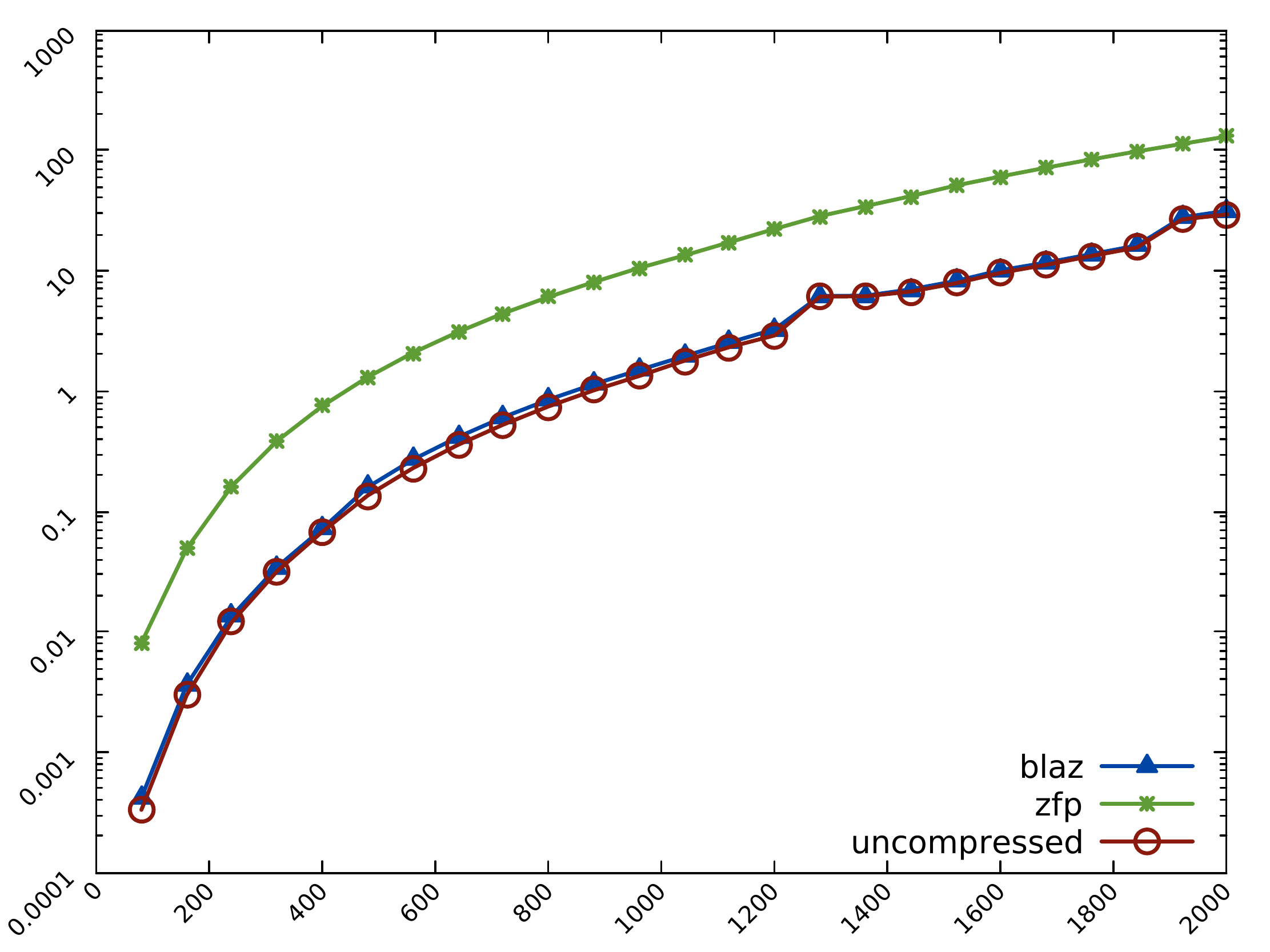}\\
    \end{tabular}
    }
    \vspace{0.2cm}
    \hrule
    \caption{\label{figtime}Time measurement of operations in function of the size of the matrices (time given in logarithmic scale). Top left: Addition. 
    Top right: Multiplication by a constant.
    Bottom left: Dot product.  Bottom right:  Matrix multiplication.}
    \end{figure*}
    
    \begin{figure*}[t]
        \hrule
        \centerline{
        \begin{tabular}{ccc}    
        \includegraphics[width = 0.31\textwidth]{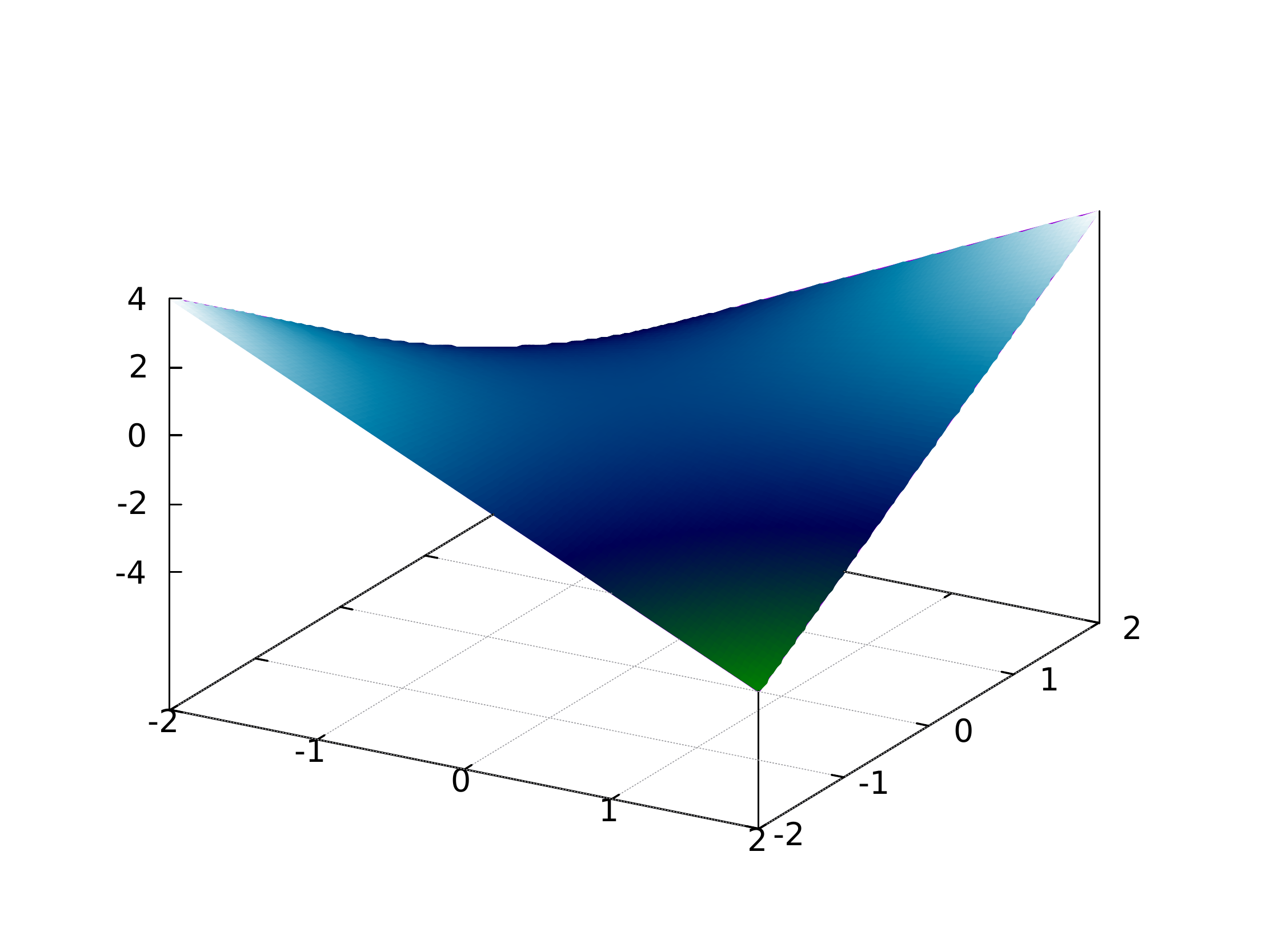} &
        \includegraphics[width = 0.31\textwidth]{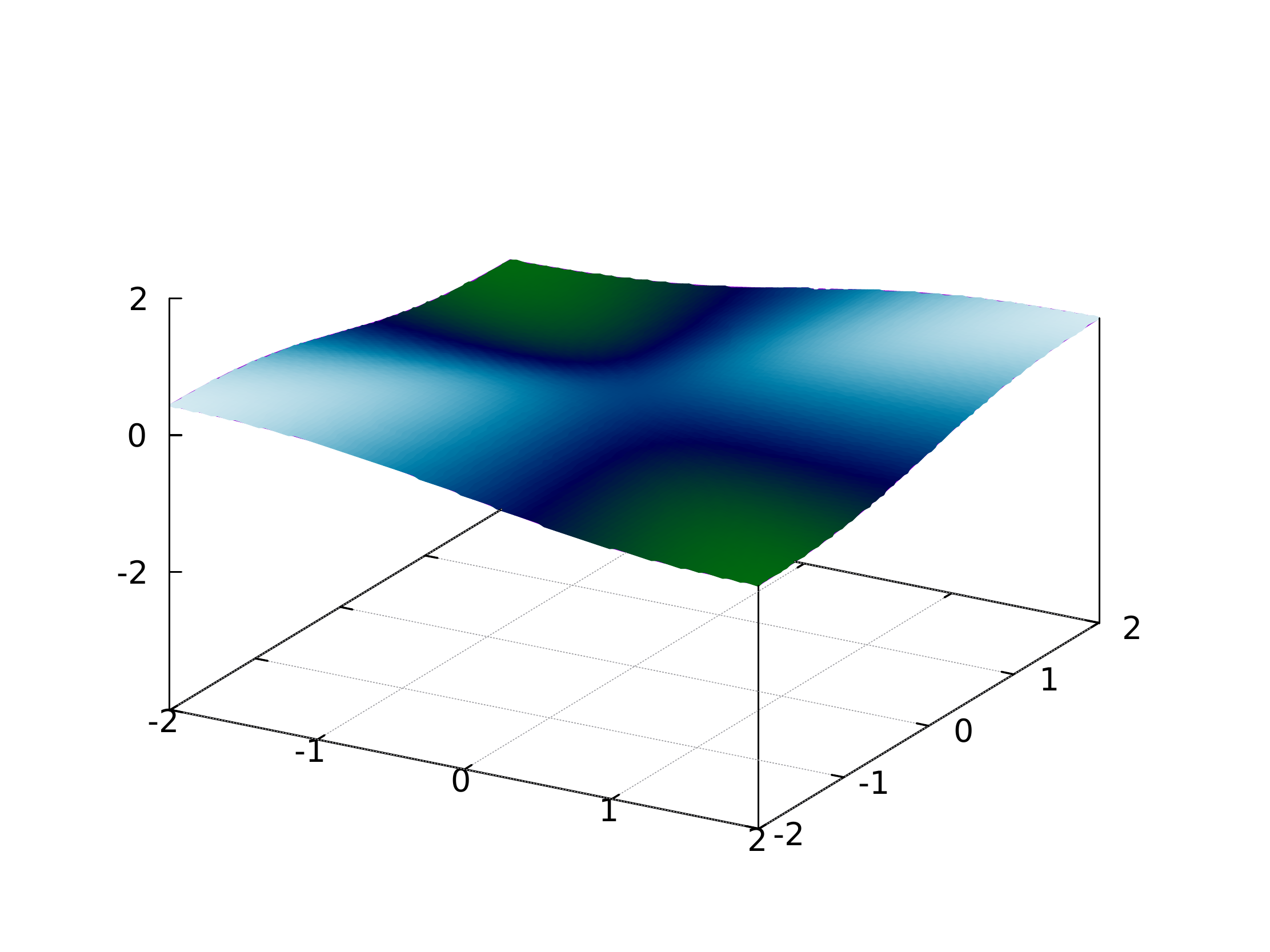} &
        \includegraphics[width = 0.31\textwidth]{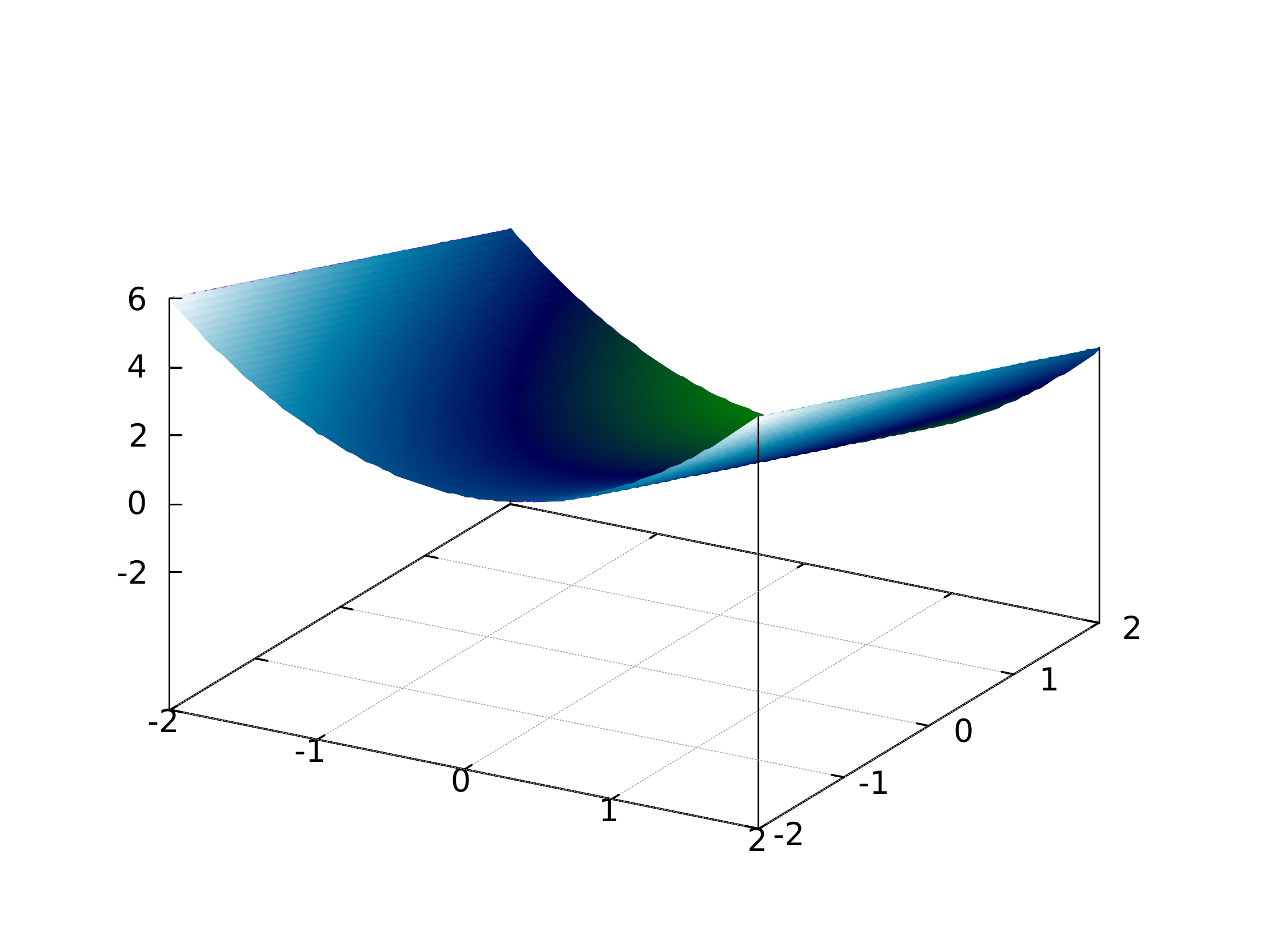} \\
        $f_1(x,y)  =x\times y$ & $ f_2(x,y)= \frac{x y}{1 + x^2 + y^2}$ & $f_3(x,y) = x^2-y$\\
        \includegraphics[width = 0.31\textwidth]{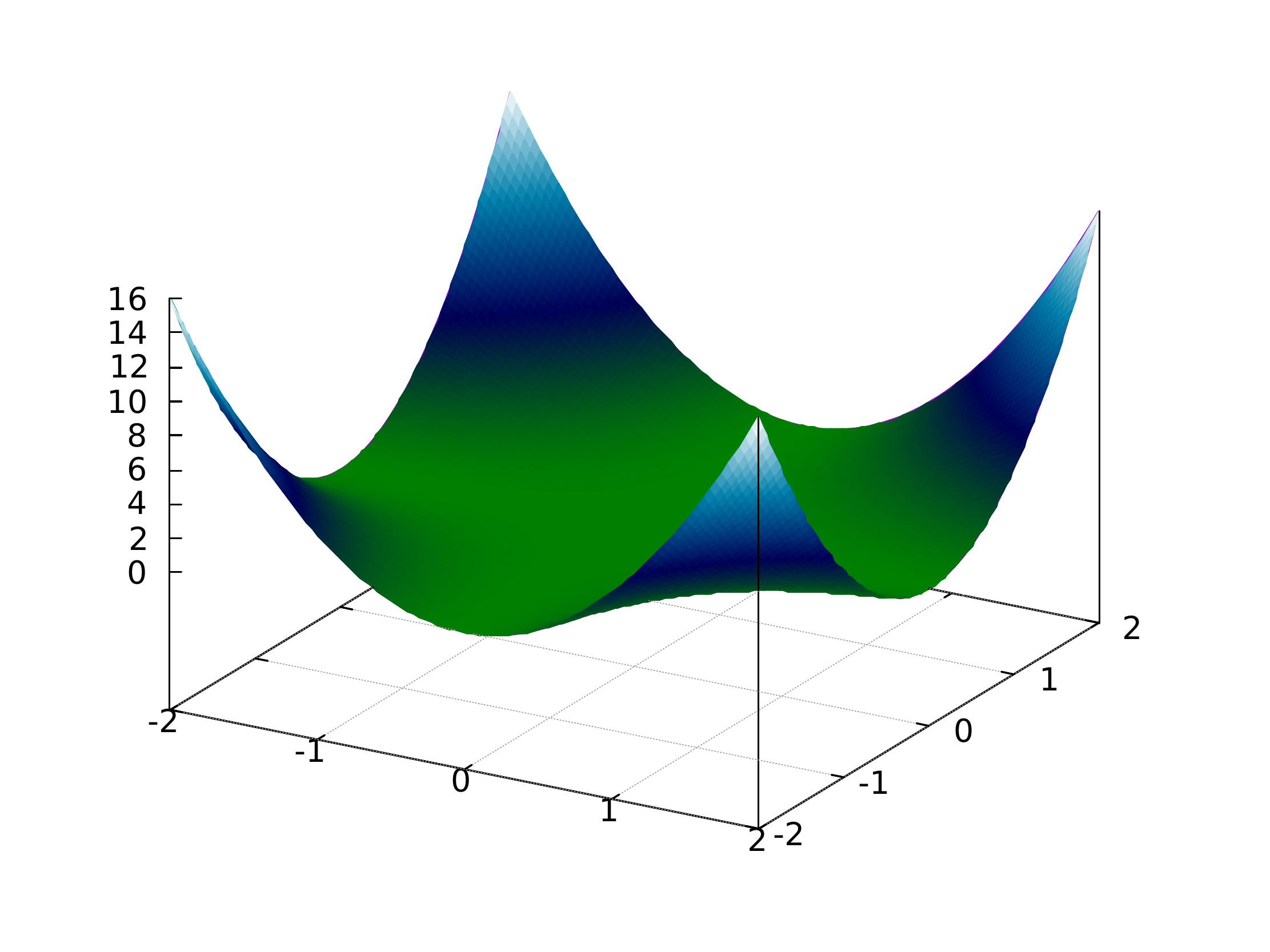} &
        \includegraphics[width = 0.31\textwidth]{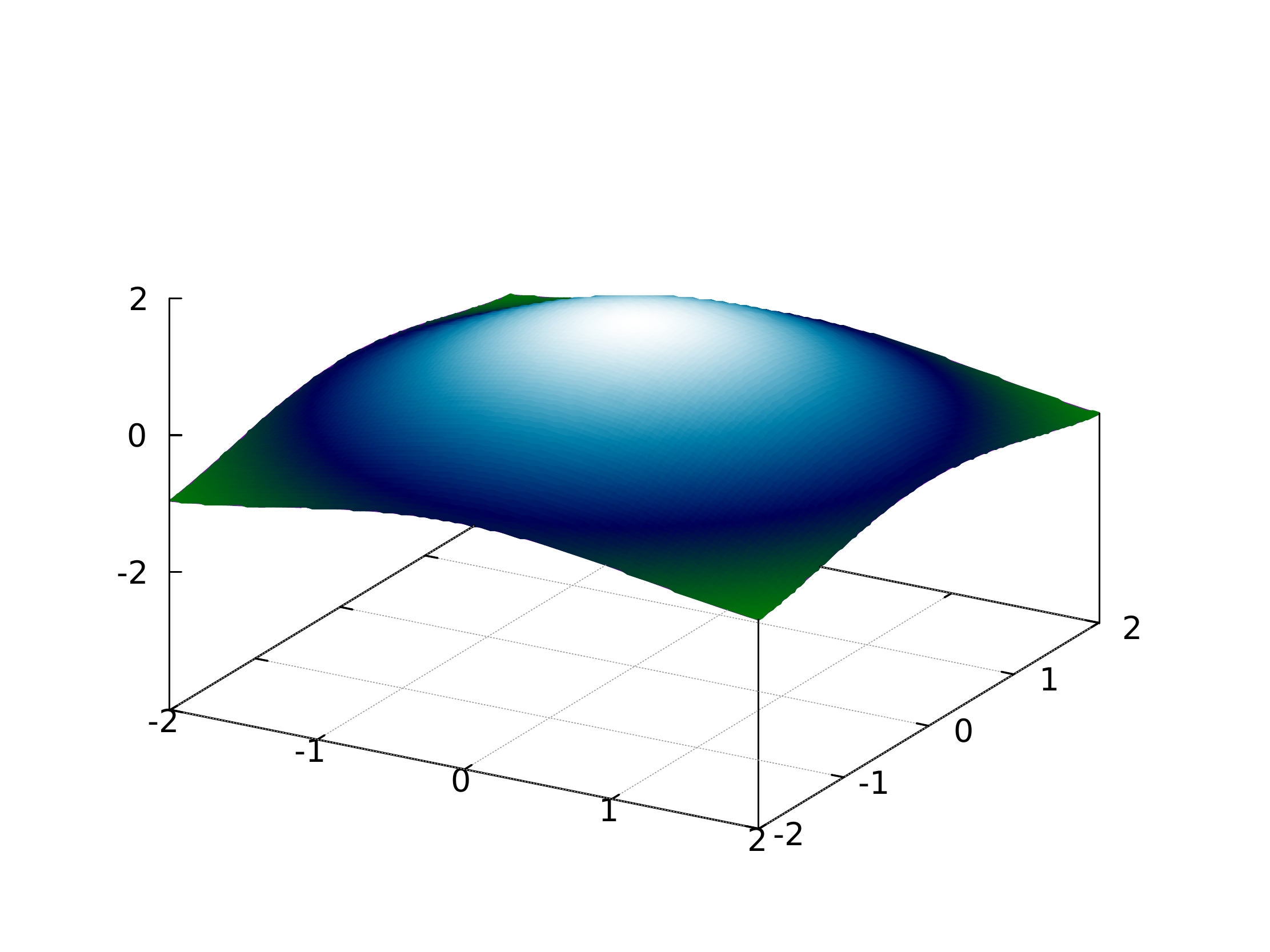} &
        \includegraphics[width = 0.31\textwidth]{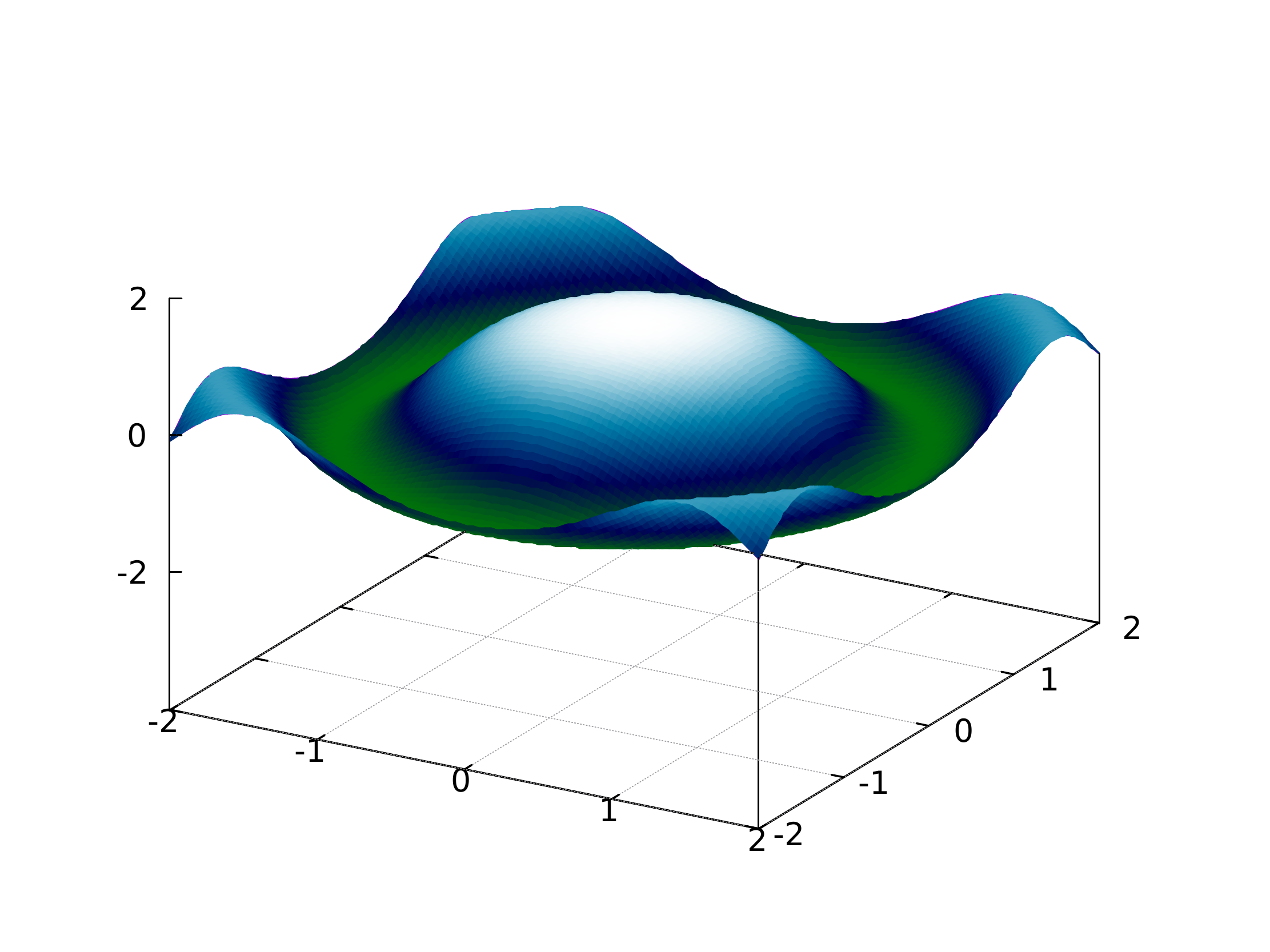}\\
        $f_4(x,y)=x^2 \times y^2$ & $f_5(x,y)=\cos(\sqrt{x^2+y^2})$ & $f_6(x,y)=\cos(x^2+y^2)\cdot \text{e}^{-0.1\cdot (x^2+y^2) }$\\
        &&\\
        \end{tabular}}
        \hrule
        \caption{\label{figcurves}Functions used to test the accuracy of basic linear algebra operations among \texttt{blaz} matrices.}
    \end{figure*}

\subsection{Performance}
\label{xpperf}

In this section, the performances of the basic linear algebra operations performed by \texttt{blaz} are compared to the same operations
done without any compression and to the case where matrices are compressed by \texttt{zfp} \cite{Lin14} (\texttt{C++} version).

First, we compare the execution time of operations among matrices compressed with \texttt{blaz} to the execution time of
the same operations performed among uncompressed matrices.
We use square matrices of sizes ranging from $80$ to $2000$ and the measures are done for addition, multiplication by a constant and for the
dot product. Our results are displayed in Figure \ref{fighs}. Let us mention that
the time in the $y$-axis of Figure \ref{fighs} is given in logarithmic scale and that the execution time for the uncompressed dot product is almost 
null and is not represented. The first noticeable point is that the
addition and multiplication by a constant are significantly faster with our compressed matrices than without any compression.
Indeed, for square matrices of size $2000$, the execution time is divided by $62$ for addition ($0.15$ms vs $9.35$ms)  and
by $58$  for the multiplication by a constant ($0.14$ms vs $8.14$ms.) Our technique makes it possible to compute faster than without any compression 
when small losses of accuracy are acceptable (see Section \ref{xpacc}.)
Concerning the dot product, the compressed operation is  slower
with a compressed matrix than without compression ($0.12$ms vs $0.004$ms which is $30$ times slower.)
This is due to the fact that a partial decompression of the matrix is done in this case (see Algorithm \ref{algodp}).

To complete our study, we have also measured the time needed by \texttt{blaz} to compress and decompress matrices. These times are given in Figure \ref{figcu}.

Second, we compare \texttt{blaz} to the well-known compressor \texttt{zfp}. For both tools, we start with compressed matrices.
In the case of \texttt{zfp}, when a matrix element is read, the corresponding block is uncompressed and the value is returned. Conversely, 
when some value is assigned to a certain element, \texttt{zfp} directly modifies and re-compress the corresponding  block. 
In the case of \texttt{blaz}, no decompression/re-compression is needed, the operations are carried out directly on the compressed matrices.
The execution times are given in Figure \ref{figtime} for our basic linear algebra operations. As in Figure \ref{fighs}, the execution time represented on the
$y$-axis is displayed in logarithmic scale.

\begin{table}[tb]\scriptsize
    \hrule
    \vspace{0.15cm}
\centerline{\begin{tabular}{cccccccc}
              & $M_1$  & $M_2$   & $M_3$   & $M_4$  & $M_5$  & $M_6$ \\
\\
              & \multicolumn{6}{c}{\textbf{Compression/Decompression of Matrices}}\\
\\
\texttt{blaz} &0.43\%  & 0.39\%  & 0.53\%  & 0.44\% & 1.95\% & 1.17\%\\
\texttt{zfp}  &0.0006\%&0.0009\% & 0.02\%  & 0.002\%& 0.13\% & 0.15\%\\
\\
& \multicolumn{6}{c}{\textbf{Additions of Compressed Matrices (\texttt{blaz} \& \texttt{zfp})}}\\
\\
$M_1$ &  $-$  & 0.98\% & 0.67\%   & 0.91\% & 0.72\% & 0.78\% \\
$M_2$ & 0.001\%& $-$     & 0.62\%  & 1.07\%  & 1.76\% & 1.61\% \\
$M_3$ &  0.82\%& 0.12\%  &  $-$    & 2.27\% & 0.71\% & 0.65\% \\
$M_4$ & 0.03\% & 0.42\%  & 0.27\%  &  $-$   & 1.68\% & 1.70\% \\  
$M_5$ & 0.68\% & 1.62\%  & 1.25\%  & 0.89\% & $-$    & 0.94\% \\ 
$M_6$ & 0.31\% & 1.27\%  & 0.25\%  & 2.32\% & 0.16\% & $-$ \\
\\
& \multicolumn{6}{c}{\textbf{Multiplications of Compressed Matrices by Constants}}\\
\\
\texttt{blaz} &0.43\%  & 0.44\%  & 0.53\%  & 0.44\% & 1.95\% & 1.17\%\\
\texttt{zfp}  &0.001\% & 0.001\% & 0.03\%  & 0.005\%& 0.12\% & 0.23\%\\
\\
& \multicolumn{6}{c}{\textbf{Multiplications of Compressed Matrices (\texttt{blaz})}}\\
\\
$M_1$ & $2.0e^{-4}$ & $2.0e^{-4}$ &$ 1.0e^{-4}$  & $5.1e^{-2}$  & $3.6e^{-2}$  & $9.2e^{-2}$ \\
$M_2$ & $2.0e^{-4}$ & $2.0e^{-4}$ &$ 1.0e^{-4}$  & $5.3e^{-2}$  & $3.6e^{-2}$  & $1.0e^{-1}$ \\
$M_3$ & $3.4e^{-2}$ & $3.5e^{-2}$ &$ 7.9e^{-3}$  & $9.0e^{-4}$  & $5.0e^{-4}$  & $5.0e^{-4}$ \\
$M_4$ & $3.7e^{-2}$ & $3.4e^{-2}$ &$ 2.1e^{-3}$  & $8.0e^{-4}$  & $6.0e^{-4}$  & $5.0e^{-4}$ \\
$M_5$ & $4.4e^{-2}$ & $4.5e^{-2}$ &$ 1.7e^{-3}$  & $8.0e^{-4}$  & $1.1e^{-3}$  & $1.7e^{-3}$ \\
$M_6$ & $1.8e^{-1}$ & $3.7e^{-1}$ &$ 2.0e^{-2}$  & $8.0e^{-4}$  & $7.0e^{-4}$  & $2.9e^{-3}$ \\
\\
\end{tabular}}
\vspace{0.15cm}
\hrule
\caption{\label{tabaccu}Relative errors on the results of operations among the matrices $M_1$ to $M_6$.
For addition: Upper right triangle: \texttt{blaz}. Lower left  triangle: \texttt{zfp}.
For multiplication: \texttt{blaz} only.
}
\end{table}
\normalsize

Again, we take square matrices of size ranging from $80$ to $2000$ and we show on a same graph the execution time taken by \texttt{blaz}, \texttt{zfp}
and without compression. The top two graphs of Figure \ref{figtime} are for addition and multiplication by a constant. The main observation
is that \texttt{zfp} introduces a huge overhead due to the decompression and re-compression of matrices. For addition, taking as reference the execution time
without any compression, \texttt{blaz} and \texttt{zfp} introduce respectively a speedup of $3$ and a speed down of $121$ for matrices of size $80$
and a speedup of $62$ and a speed down $115$ for matrices of size $2000$. In other terms, the addition is more than $50$ times faster with \texttt{blaz} and more than
$100$ times slower with \texttt{zfp} compared to the uncompressed operation. Note that similar results are observed for the multiplication by a constant.

The execution times for the dot product are displayed in the left bottom corner of Figure \ref{figtime}.
As already mentioned, this operation carried out with \texttt{blaz} matrices is slower than without any compression. Nevertheless, it remains
significantly faster than with \texttt{zfp} matrices. Again, taking as reference the execution time without compression, \texttt{blaz} is $31$ times slower for 
matrices of size $2000$ while \texttt{zfp} is $62$ times slower. 
In this case, \texttt{blaz} remains  $2$ times faster than \texttt{zfp}.

Finally, the graph in the bottom right corner of Figure \ref{figtime} show the performances of matrix multiplication. A first observation
is that \texttt{blaz} is not much slower than the uncompressed operation. This is due to the fact that time spent in the matrix multiplication
dominates the partial compression/decompression time. The second observation is that \texttt{zfp} is much slower than \texttt{blaz} (approximately $10$ times.)

\subsection{Accuracy Measurements}
\label{xpacc}

In this section, we introduce a second set of experiments concerning the accuracy of \texttt{blaz}. Again, we compare our tool to \texttt{zfp}.
Contrarily to the execution time measurements introduced in Section \ref{xpperf}, the values of the matrix elements now matter.
We use matrices corresponding to the discretisation of the six non-linear functions $f_k\ : \mathbb{R}^2\rightarrow \mathbb{R}, \ 1\le k\le 6$, 
presented in Figure \ref{figcurves}. We set $M_{k_{ij}}=f_k(x_j,y_i)$ where the points $(x_j,y_i)$ are taken in the range $[-2,2]\times[-2,2]$ with
a constant step depending on the size of $M_k$.

Our experimental results are displayed in Table \ref{tabaccu}.
They correspond to mean relative errors computed as follows. Let $M$, $M_1$ and $M_2$ denote square matrices of size $n$ whose elements are 
floating-point numbers. Let  $\mathcal{C}$ and
$\mathcal{U}$ denote the compression and decompression functions and let $M'= \mathcal{U}(\mathcal{C}(M))$, the mean relative error $e$ on $M'$ is
\begin{equation}\small\label{eqmean}
e = \frac{1}{n^2}\sum_{i,j=1}^n \left| \frac{M'_{ij}-M_{ij}}{M_{ij}}\right| \enspace .
\end{equation}
We proceed similarly for matrices resulting from basic linear algebra operations. For instance, we apply Equation (\ref{eqmean}) to $M = M_1+M_2$ and
$M' =  \mathcal{U}(\mathcal{C}(M_1)+ \mathcal{C}(M_2))$.

The first two lines of Table \ref{tabaccu} indicate the mean relative error on the compression/decompression of the matrices $M_1$ to $M_6$ using
\texttt{blaz} and \texttt{zfp}. While \texttt{zfp} is more accurate than \texttt{blaz}, this latter remains accurate with relative errors of less than
$0.5\%$  in half of the cases and never greater than $2\%$.
  
The second part of Table \ref{tabaccu} is dedicated to addition. We display the mean relative errors for all the additions $M_i+M_j$, $1\le i,j\le 6$, 
$i\not= j$ performed on \texttt{blaz} matrices (upper right triangle) and on \texttt{zfp} matrices (lower left triangle).
For example, the mean  relative errors on $M_1+M_6$ with \texttt{blaz} and \texttt{zfp} respectively are $0.78\%$ and $0.31\%$. The errors
introduced by \texttt{blaz} and \texttt{zfp} are comparable (same magnitude in general) even if \texttt{zfp} is better in most cases (at the price of the
huge overhead in terms of execution time shown in Section \ref{xpperf}.) In some cases, \texttt{blaz} is more accurate than \texttt{zfp} (e.g. for $M_4+M_6$
\texttt{blaz} error is $1.70\%$ while \texttt{zfp} error is $2.32\%$.) We may conclude that, for addition, there is not a clear advantage to use
\texttt{zfp} instead of \texttt{blaz}.

The third part of Table \ref{tabaccu} deals with the multiplication by a constant and, in our experiments, the matrices $M_1$ to $M_6$ are multiplied by $2$. 
While being less accurate than \texttt{zfp}, the accuracy of the computations remain accurate, with relative errors around $1\%$.
Note that changing the constant does not change the relative error.

The last part of Table \ref{tabaccu} is for matrix multiplication. Let us mention that, in \texttt{blaz}, the value and accuracy of the  dot product 
$\langle M,M'\rangle_{ij}$ is the same than these of the element $(i,j)$ of the matrix $M\times M'$. Then we only discuss the accuracy of the
matrix multiplication which encompasses the one of the dot product.
The matrix product not being symmetric in general, we display in Table \ref{tabaccu} the mean accuracies of all the products $M_i\times M_j$,
$1\le i,j\le 6$, computed following Equation (\ref{eqmean}). 
These accuracies are all for \texttt{blaz}. The errors obtained with \texttt{zfp} being almost zero in any case, we do not display them.
We can see that the accuracy of \texttt{blaz} for matrix multiplication is very good (relative errors of order $10^{-3}$ or less in more than half of the cases
and never greater than $10^{-1}$ in magnitude), even if they remain worse than \texttt{zfp}. 
We can also remark that \texttt{blaz} multiplication is more accurate than \texttt{blaz} addition. This is due to the fact that a partial
decompression is done for the multiplication, contrarily to the case of addition. This better accuracy then results from more computations and,
in future work, we would like to design even more frugal algorithms (yet less accurate) for the dot product and multiplication.

%% file: cc.tex
\section{Related Work}

Roughly speaking, compressors for scientific data (i.e. arrays of floating-point numbers) can be classified into two categories: 
Either the user sets the compression rate and this determines the accuracy of the encoding or
the user sets the accuracy and this determines the compression rate. The compressors \texttt{zfp} \cite{Lin14} and
\texttt{fpzip} \cite{LI06} before it 
have been initially designed as compressors of the first category. Conversely, \texttt{sz} \cite{DC16,Lal18} has been designed as a compressor 
of the second category. Elements of comparison between these two approaches can be found in \cite{Tal19}.
Note that the current versions of \texttt{zfp} and \texttt{sz} offer many options which make them encompass the categories mentioned earlier.
Let us also mention an extension of \texttt{sz} \cite{Zal21} which uses spline functions for more accuracy 
and another approach based on topological control \cite{SPCT18}. 

Another related topic is about the use of GPUs for scientific data compression  \cite{JGBPTTA20,KTF21,Tal20,Tal21}.  
Finally, much attention has also been paid recently to the compression of neural networks \cite{JDLTTC19,JGMGG20,ON20}. We strongly believe that
the ideas developed in this article could be useful in the context of neural networks which perform mainly basic linear algebra operations.

\section{Perspectives and Conclusion}
\label{seccc}

The work presented in this article opens a new research direction and many perspectives remain to be explored. 
First of all, we would like to determine formal error bounds for our compressor (in the spirit of \cite{DFHS19}) and, more interestingly, for the operations
on compressed matrices. We strongly believe that the error on the result of some operation between compressed matrices
can be expressed in function of the errors on the compressed matrices corresponding to the operands.

Second, we aim at making the compression rate tunable, more or less compression being permitted with the corresponding impact on accuracy.
This can be achieved by skewing the quantization or by introduction an interpolation stage in our scheme. 
For example, we could add a Lorenzo predictor which would be compatible with our basic linear algebra operations~\cite{ILal03}.

Third, we believe that compressors allowing to compute on the compressed matrices could be useful in other contexts.
For example a compressor for integers could be useful to perform fast and frugal basic linear algebra operations in the context
of image processing. As well, a \texttt{blaz}-like compressor designed for embedded systems could be useful for many applications
with memory constraints. 
 
Next, we plan to extend our library, with new linear algebra operators but also,
our compression scheme processing every $8\times 8$ block of a matrix independently, we would like to develop a parallel
implementation of our \texttt{blaz} library, typically based on the MPI library \cite{Gal14}.

Yet another perspective is to use compressed matrices in precision tuning tools \cite{CA20}. Precision tuning consists of reducing
the precision of the variables of a program up to some accuracy requirement set by the user. It would be interesting to
use the precision recommended by the precision tuning tools to compress arrays up to this requirement.

Finally, we aim at testing further the \texttt{blaz} library. We aim at evaluating it in the context of 
large numerical simulations to assess its performances in term of execution-time and accuracy on more realistic
use-cases \cite{Cal19,Zal20}.